\documentclass[10pt,twocolumn, a4paper]{IEEEtran}

\usepackage{epsfig}
\usepackage{amsmath}
\usepackage{latexsym}
\usepackage{amsfonts,amsmath,color,amssymb,amsxtra, graphicx, times}
\usepackage[ansinew]{inputenc}
\usepackage{subfigure}

\date{}
\begin{document}
\newcommand{\edgesin}{\Gamma_{I}(v)}
\newcommand{\edgesout}{\Gamma_{O}(v)}
\newcommand{\indegree}{\delta_{I}(v)}
\newcommand{\outdegree}{\delta_{O}(v)}
\newcommand{\field}{{\mathbb F}_{q}}
\newcommand{\comment}[1]{}
\newcommand{\fig}[1]{{\itshape Figure~\ref{#1}}}
\newcommand{\alg}[1]{{\itshape Algorithm~\ref{#1}}}
\newcommand{\sect}[1]{{\itshape Section~\ref{#1}}}
\newcommand{\tabl}[1]{{\itshape Table~\ref{#1}}}
\newcommand{\eq}[1]{\eqref{#1}}
\newcommand{\secref}[1]{Section~\ref{#1}}
\newcommand{\ie}{i.e.~}
\newcommand{\eg}{e.g.~}
\newcommand{\theorref}[1]{{\itshape Theorem~\ref{#1}}}
\newcommand{\proprref}[1]{{\itshape Proposition~\ref{#1}}}
\newcommand{\lemmarref}[1]{{\itshape Lemma~\ref{#1}}}
\newcommand{\boxend}{\hfill$\blacksquare$} 
\newtheorem{theorem}{Theorem}
\newtheorem{lemma}{Lemma}
\newtheorem{proposition}{Proposition}
\newtheorem{corollary}{Corollary}
\newtheorem{example}{Example}
\newtheorem{definition}{Definition}
\newtheorem{remark}{Remark}

\title{Random Linear Network Coding: \\ A free cipher?}

\author{Lu\'isa Lima \qquad Muriel M\'edard \qquad Jo\~{a}o Barros
\thanks{L. Lima (luisalima@ieee.org) and J. Barros (barros@dcc.fc.up.pt) are with the Instituto de Telecomunica\c{c}\~oes (IT) and the
Department of Computer Science, Faculdade de Ci\^{e}ncias da
Universidade do Porto, Portugal.
M. M\'edard (medard@mit.edu) is with the Laboratory for Information and Decision Systems at the Massachusetts Institute of Technology.
This work was partly supported by the Funda\c{c}\~{a}o
para a Ci\^{e}ncia e Tecnologia
(Portuguese Foundation for Science and Technology)
under grant SFRH/BD/24718/2005 and by AFOSR under grant "Robust Self-Authenticating Network Coding" AFOSR 000106.
 Part of this work was done while the first author was
a visiting student at the Laboratory for Information and Decision Systems at the Massachusetts Institute of Technology.
}
}

\maketitle

\begin{abstract}
We consider the level of information security provided by
random linear network coding in network scenarios in which all nodes
comply with the communication protocols yet are assumed
to be potential eavesdroppers (i.e.~``nice but curious").
For this setup, which differs from  wiretapping
scenarios considered previously, we develop a natural {\it algebraic security}
criterion, and prove several of its key properties.
A preliminary analysis of the impact of
 network topology on the overall network coding security,
 in particular for complete directed acyclic graphs, is also included.
\end{abstract}


\section{Introduction}\label{sect:Introduction}

Under the classical networking paradigm, in which intermediate nodes are only allowed to store and forward packets,
information security is usually viewed as an independent feature with little or no relation to other communication tasks.
In fact, since intermediate nodes receive exact copies of the sent packets, data confidentiality is commonly ensured by
cryptographic means at higher layers of the protocol stack. Breaking with the ruling paradigm,
network coding allows intermediate nodes to mix information from different data flows
~\cite{ahlswede2000nif, koetter2003aan} and thus provides an intrinsic level of data security
--- arguably one of the least well understood benefits of network coding.

\par Previous work on this issue has been mostly concerned with constructing codes capable of spliting the data among different links, such that reconstruction by a wiretapper is either very difficult or impossible. In  \cite{cai2002snc}, the authors present a secure linear network code that achieves perfect secrecy against an attacker with access to a limited number of links. A similar problem is considered in \cite{feldman2004csn}, featuring a random coding approach in which only the input vector is modified. \cite{bhattad2005wsn} introduces a different information-theoretic security model, in which a system is deemed to be secure if an eavesdropper is unable to get any
decoded or decodable (also called {\it meaningful}) source data. Still focusing on wiretapping attacks, \cite{jain2004sbn} provides a simple security protocol exploiting the network topology: an attacker is shown to be unable to get any meaningful information unless it can access those links that are necessary for the communication between the legitimate sender and the receiver, who are assumed to be using network coding.
As a distributed capacity-achieving approach for the multicast case, randomized network coding \cite{ho2003bco,ho2003rnc} has been
shown to extend naturally to packet networks with losses \cite{lun2005crc} and
Byzantine modifications (both detection and correction \cite{ho2004bmd,jaggi2005cae,jaggi2006rnc,jaggiThesis}).
~\cite{tan2006snc} adds a cost criterion to the secure network coding problem, providing heuristic solutions for a coding scheme that minimizes both the network cost and the probability that the wiretapper is able to retrieve all the messages of interest.

\begin{figure}[t!]
\centering
\includegraphics[height=4cm]{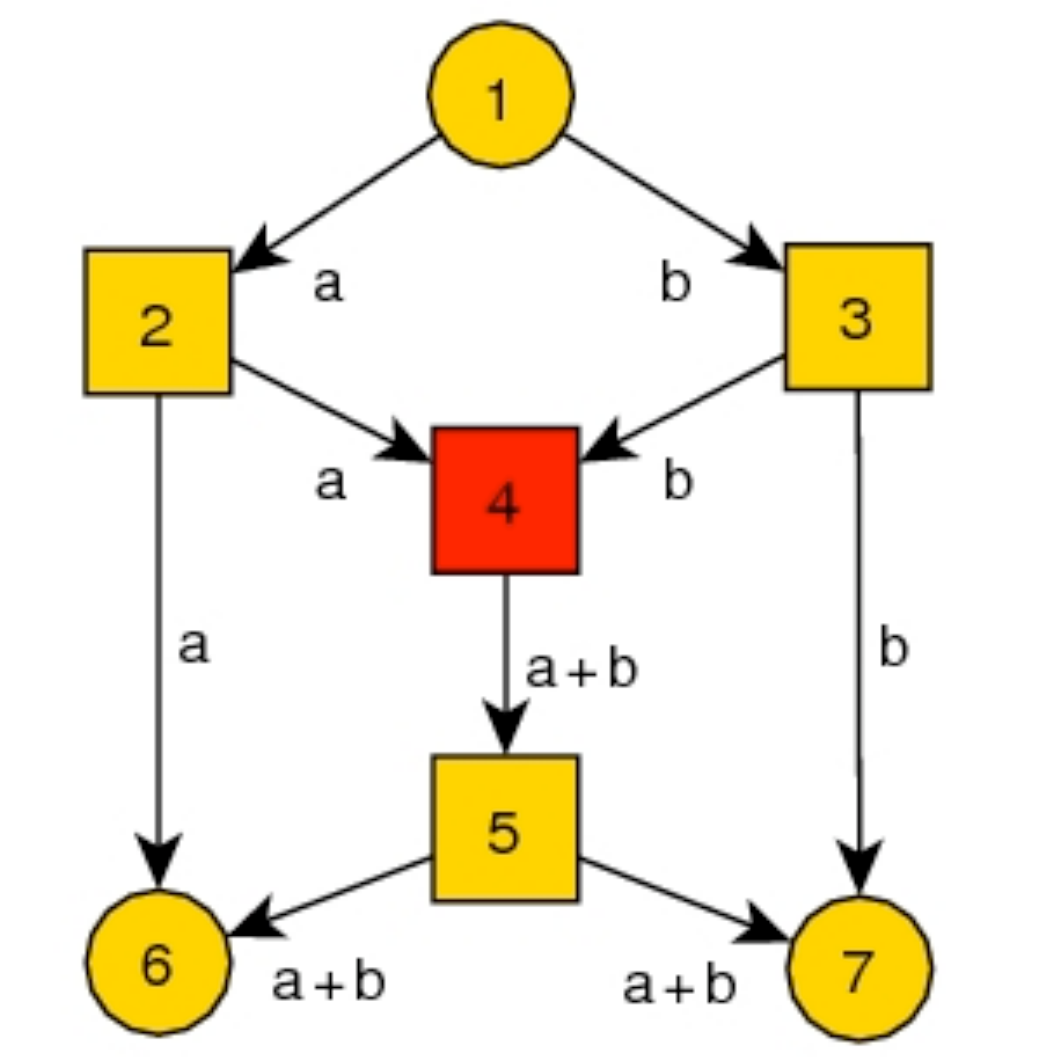}
\caption{Canonical Network Coding Example. In this image, intermediate nodes are represented with squares. With this code, node 4 is a vulnerability for the network since it can decode all the information sent through it. Note that the complete opposite happens for node 5, that receives no meaningful information whatsoever.
}
\label{fig:BU}
\end{figure}
In this work, we approach network coding security from a different angle: our focus is {\it not} on the threat posed by external wiretappers but on the more general threat posed by intermediate nodes. We assume that the network consists entirely of ``nice but curious" nodes, i.e. they comply with the communication protocols (in that sense, they are well-behaved) but may try to acquire as much information as possible from the data that passes through them (in which case, they are potentially malicious). This notion is highlighted in the following example.

\begin{example}
Consider the canonical network coding example with $7$ nodes, shown in~\fig{fig:BU}. Node $1$ sends a flow to sinks $6$ and $7$ through intermediate nodes $2$, $3$, $4$ and $5$. From the point of security, we can distinguish between three types of intermediate nodes in this setting: (1) those that only get a non-meaningful part of the information, such as node $5$; (2) those that obtain all of the information, such as node $4$; and (3) those that get partial yet meaningful information, such as nodes $2$ and $3$. Although this network code could be considered {\em secure} against single-edge external wiretapping --- \ie, the wiretapper is not able to retrieve the whole data simply by eavesdropping on a single edge  --- it is clearly insecure against internal eavesdropping by an intermediate node.
\end{example}

Motivated by this example, we set out to investigate the security potential of network coding.
Our main contributions are as follows:
 \begin{itemize}
 \item {\it Problem Formulation}: We formulate a secure network coding problem, in which all intermediate nodes are viewed as potential eavesdroppers and the goal is to characterize the intrinsic level of security provided by random linear network coding.
 \item {\it Algebraic Security Criterion}: Based on the notion that the number of decodable bits available to each intermediate node is limited by the degrees of freedom it receives, we are able to provide a natural secrecy constraint for network coding and to prove some of its most fundamental properties.
 \item {\it Security Analysis for Complete Directed Acyclic Graphs}: As a preliminary step towards understanding the interplay between network topology and security against eavesdropping nodes, we present a rigorous characterization of the achievable level of algebraic security for this class of complete graphs.
 \end{itemize}
 The remainder of this paper is organized as follows. First, a formal problem statement is
 in \sect{sect:ProblemFormulation}, followed by a detailed analysis of the algebraic security
 of Randomized Linear Network Coding in \sect{sect:secRLNC}. In \sect{sect:DAG}, this analysis is carried out
 specifically for complete directed acyclic graphs.
The paper concludes with \sect{sect:ConcludingRemarks}.

\section{Problem Setup}\label{sect:ProblemFormulation}

We adopt the network model of ~\cite{koetter2003aan}: we represent the network as an acyclic directed graph $G = (V, E)$, where $V$ is the set of nodes and $E$ is the set of edges.
Edges are denoted by round brackets $e=(v,v') \in E$, in which $v=\textrm{head}(e)$ and $v'=\textrm{tail}(e)$. The set of edges that end at a vertex $v \in V$ is denoted by $\edgesin = \{ e \in E: \textrm{head}(e) = v \} $, and the in-degree of the vertex is $\indegree = |\edgesin|$; similarly, the set of edges originating at a vertex $v \in V$ is denoted by $\Gamma_{O}(v) = \{ e \in E: \textrm{tail}(e) = v \} $, the out-degree being represented by $\outdegree = |\edgesout|$.

Discrete random processes $X_{1}, ... X_{K}$ are observable at one {or more} source nodes. To simplify the analysis, we shall consider that each network link is free of delays and that there are no losses. Moreover, the capacity of each link is one bit per unit time, and the random processes $X_{i}$ have a constant entropy rate of one bit per unit time. Edges with larger capacities are modelled as parallel edges and sources of larger entropy rate are modelled as multiple sources at the same node.
{ We shall consider multicast connections as it is the most general type of single
connection; there are $d\ge1$ receiver nodes. The objective is to transmit all the source processes to each of the receiver nodes.}

In linear network coding, edge $e=(v,u)$ carries the process $Y(e)$, which is defined below:

$$Y(e) = \sum_{l: X_{l} \textrm{ generated at v}}\alpha_{l,e}X(v,l)+\sum_{e':head(e')=tail(e)}\beta_{e',e}Y(e')$$

The {\em transfer matrix $M$} describes the relationship between an input vector $\underline{x}$ and an output vector $\underline{z}$, $\underline{z} = \underline{x}M$; $M=A(I-F)^{-1}B^{T}$, where $A$ and $B$ represent, respectively, the linear mixings of the input vector and of the output vector, and have sizes $K\times|E|$ and $\nu\times|E|$. $F$ is the adjacency matrix of the directed labelled line graph corresponding to the graph $G$.
In this paper we shall not consider matrix $B$, which only refers to the decoding at the receivers. Thus, we shall mainly analyse parts of the matrix $AG$, such that $G=(I-F)^{-1}$; $\underline{a}_{i}$ and $\underline{c}_{i}$ denote column $i$ of $A$ and $AG$, respectively. 
We define the {\em partial transfer matrix} $M'_{\edgesin}$ (also called {\em auxiliary encoding vector}~\cite{lun2005crc}) as the observable matrix at a given node $v$, \ie the observed matrix formed by the symbols received at a node $v$. This is equivalent to the fraction of the data that an intermediate node has access to in a multicast transmission.

Regarding the coding scheme, we consider the random linear network coding scheme introduced in ~\cite{ho2003bco}: and thus each coefficient of the matrices described above is chosen independently and uniformly over all elements of a finite field $\field$, $q=2^m$.

Our goal is to evaluate the {\em intrinsic security} of random linear network coding, in multicast scenarios where all the intermediate nodes in the network are potentially malicious eavesdroppers.
Specifically our threat model assumes that intermediate nodes perform the coding operations as outlined above, and will try to decode as much data as possible.

\section{Algebraic Security of Random Linear Network Coding}\label{sect:secRLNC}

\subsection{Algebraic security}

\begin{figure}[h!]
\centering
\includegraphics[width=6cm]{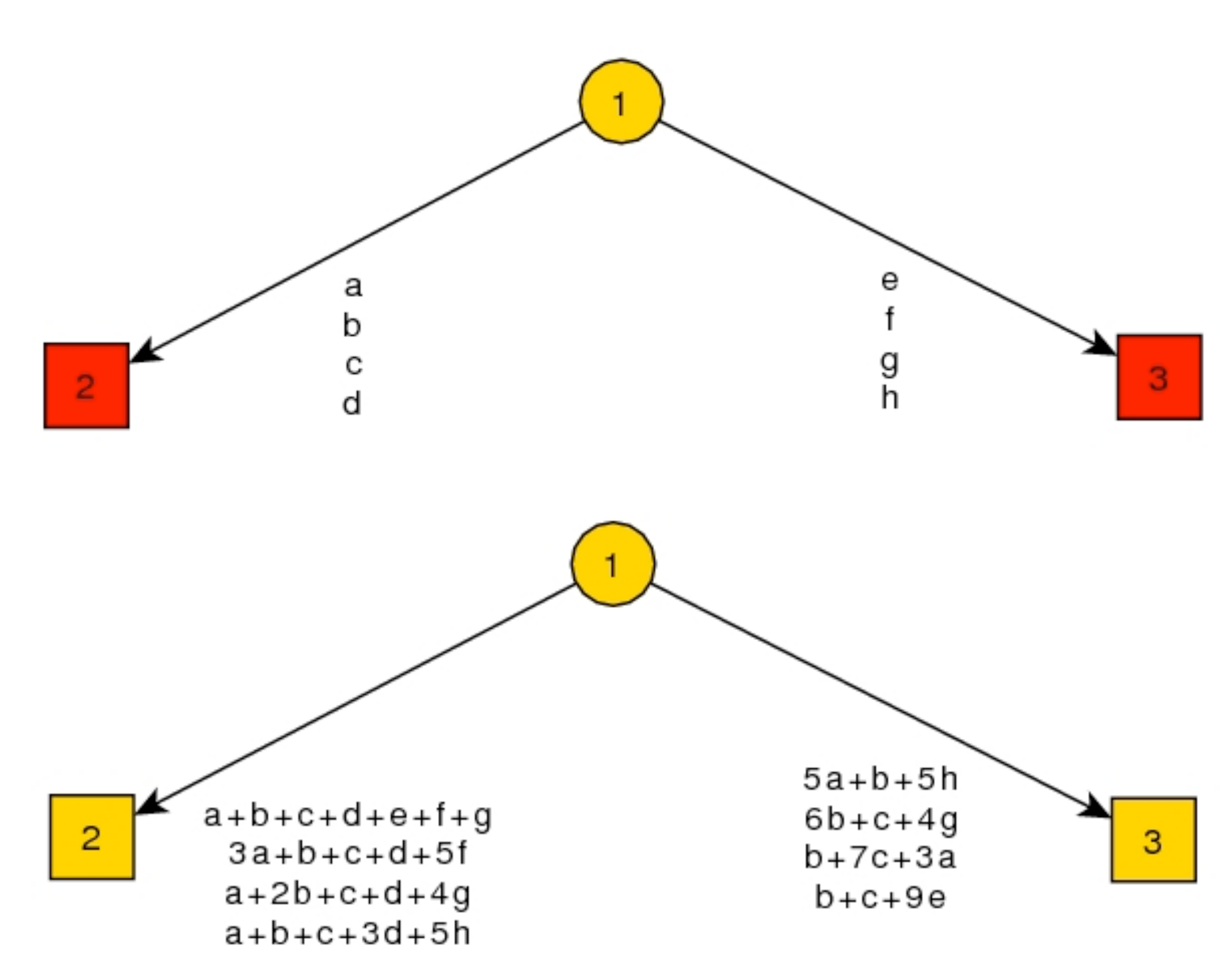}
\caption{Example of algebraic security. In the upper scheme data is not protected, whereas in the lower scheme nodes 2 and 3 are
unable to recover any data symbols.}
\label{fig:SC1}
\end{figure}

The Shannon criterion for information-theoretic security~\cite{shannon1949cta} corresponds in general terms to a zero mutual information between the cypher-text ($C$) and the original message ($M$),~\ie $I(M;C)=0$.
This condition implies that an attacker must guess $\leq H(M)$ symbols to be able to compromise the data.
With network coding, on the other hand,
 if the attacker is capable of guessing $M$ symbols,  $K-M$ additional observed symbols are required for decoding --- by noting that each received symbol is a linear combination of the $K$ message symbols from the source, we can see that
a receiver must receive $K$ coded symbols in order to recover one message symbol. Thus, as will be shown later, restricted rank sets of individual symbols do not translate
into immediately decodable data with high probability.
This notion is illustrated in \fig{fig:SC1}. In the scheme shown on top, each intermediate node can recover half of the
transmitted symbols, whereas in the bottom scheme none of the nodes can recover any portion of the sent data.

\definition[Algebraic Security Criterion]{The level of security provided by random linear network coding is measured by the
 number of symbols that an intermediate node $v$ has to guess in order to decode {\it one} of the transmitted symbols.
 From a formal point of view,
$$\Delta_{S}(v)= \frac{K - (\textrm{rank}( M'_{\edgesin}) + l_{d}}{K}, $$
  where $l_{d}$ represents the number of partially diagonalizable lines of the matrix (i.e.~the number of message symbols that can be recovered by Gaussian elimination).}

Notice that the previous definition is equivalent to computing the difference between the global rank of the code and the local rank in each intermediate node $v$. Moreover, as more and more symbols become compromised of security criteria, the level of security tends to $0$, since as we shall show in this section, with high probability the number of individually decodable symbols $l_{d}$ goes to zero
as the size of the field goes to infinity.

\subsection{Security Characterization}

We are now ready to solve the problem of characterizing the algebraic security of random linear network coding. The key to our proofs is to analyze the properties of the partial transfer matrix at each intermediate node. 
Recall that there are two cases in which the intermediate node can gain access to relevant information: (1) when the partial transfer matrix has full rank and (2) when the partial transfer matrix has diagonalizable parts. Thus, we shall carry out independent analyzes in terms of rank and in terms of partially diagonalizable matrices.

The following lemmas will be useful.
\lemma{In the random linear network coding scheme, $$ P(\Delta_{S}>0) \leq P(\exists v: \indegree > K). $$ }\label{prop:rank}
\begin{proof} See the {\it Appendix}. \end{proof}
It follows from this lemma that it is only necessary to consider the case in which  $K \leq \indegree $.

\begin{lemma}
The probability that a linear combination of independent and uniformly distributed values in $\field$
yields the zero result is bounded by $$P(X_{lin}=0) \leq \frac{2q+h(q)}{q^2},$$ where $h(q)$ is a function
such that $O(h(q))<O(q^2)$. Moreover, $P(X_{lin}=0)$ tends to $0$ when $q\rightarrow\infty$.
\label{prop:linCombFq}
\end{lemma}
\begin{proof} See the {\it Appendix}. \end{proof}

\lemma{The probability of obtaining $y$ zeros in one line of the $\xi\times\xi$ transfer matrix $M$ is bounded by
$$P(Y=y) \leq \dbinom{\xi}{\xi-y}\left(\frac{2q+h(q)}{q^2}\right)^{y}\left(1-\frac{2q+h(q)}{q^2}\right)^{\xi-y}.$$
}\label{prop:lineTransferMatrix}
\begin{proof} See the {\it Appendix}. \end{proof}

\theorem{Let $P(l_{d}>0)$ be the probability  of recovering a strictly positive number of symbols $l_{d}$ at the intermediate nodes  with $\indegree \leq K-1$ by Gaussian elimination. Then, $P(l_{d}>0)\rightarrow0$ with $q\rightarrow \infty$ and $K\rightarrow \infty$.}\label{theorem:diag}

\begin{proof}
Let $M'$ be the transpose of the partial transfer matrix at some vertex $v$, $M'=M_{\edgesin}^T$. We consider the process of Gaussian elimination of $M'$.
It is unnecessary to consider rank $K$, since in that case the matrix, w.h.p, is invertible and hence diagonalizable~\cite{ho2003rnc}. Thus, $M'$ is a
$\indegree\times K$ matrix, $\indegree < K$.

We prove the theorem constructively by analysing the probability of having $K-1$ zeros in one or more lines of $M'$. Let $p$ be the probability of having $K-1$ zeros in a line of $M'$, and let $X$ be a random variable representing the recoverable number of symbols when an intermediate node has $\indegree$ degrees of freedom. It follows from \lemmarref{prop:lineTransferMatrix} that

$$p=\dbinom{K}{K-1}\left(\frac{2q+h(q)}{q^2}\right)^{1}\left(1-\frac{2q+h(q)}{q^2}\right)^{K-1}.$$

In the base case with $\indegree=1$, at most $X=1$ symbols can be recovered, since there are not enough degrees of freedom to perform Gaussian elimination and the only chance for recovering a symbol is that the line of the matrix $M$ already has $K-1$ zeros. The probability for this is $p$.

In the case that $1<\indegree<K$, we can obtain directly a number $L=l$ of lines with $K-1$ zeros, and a number $\indegree-l$ of lines in the opposite situation. Since we have $\indegree$ degrees of freedom to perform Gaussian elimination, we can obtain at most $\indegree$ symbols by successive elimination. At each step the probability of obtaining a line with $K-1$ zeros is bounded by $p$.

By analysing the different possibilities of combinations for the lines that already have $K-1$ zeros and the ones that can be obtained by Gaussian elimination, we get
$$P(X=x) \leq \sum_{l=0}^{x}\dbinom{\indegree}{l}p^l(1-p)^{\indegree-l}P_{l}(X=x)$$
$$P_{l}(X=x) \leq \dbinom{\indegree-l}{x-(\indegree-l)}p^{x-\indegree+l}(1-p)^{2\indegree-2l-x},$$
where $P_{l}(X=x)$ represents $P(X=x|L=l)$.

Approximating the binomial distribution by a normal distribution yields
\begin{align*}
P_{l}(X=x) \approx \frac{e'}{\sqrt{2\pi(\indegree-l)p(1-p)}},
\end{align*}
where
\begin{align*}
e'=\exp\left(-\frac{1}{2}\frac{(x-(\indegree-l)p)^2}{(\indegree-l)p(1-p)}\right)
\end{align*}
Since $p\rightarrow p*<1$, we can state that, when $q\rightarrow \infty$ and $p\rightarrow 0$ is $\approx \exp(x^2)$. When $K$ goes to $\infty$, so does $x$, and hence
$$\exp(x^2)_{x\rightarrow\infty}\rightarrow 0,$$
and
$$P_{l}(X=K-1)_{q\rightarrow\infty, K\rightarrow\infty}\rightarrow0.$$
Since
$$P(X=K-1) = \sum_{l=0}^{K-1}\dbinom{\indegree}{l}p^l(1-p)^{\indegree-l}P_{l}(X=K-1),$$
and $P_{l}(X=K-1)$ decreases exponentially, and $l$ only increases linearly,
$$P(X=K-1)_{q\rightarrow 0, K\rightarrow \infty}\rightarrow0.$$
The probability of obtaining $X < K-1$ symbols is bounded by $P(X=K-1)$; it follows that the probability of decoding $X$ symbols
with any $\indegree<K$ goes to zero as $q$ and $K$ tend to infinity.
\end{proof}


\section{Algebraic Security of the Complete Graph}\label{sect:DAG}

Notice that, in consequence of the property outlined in Lemma ~\ref{prop:rank}, the algebraic security of a graph is topology dependent. A node with $\indegree\ge K$  will not necessarily receive a full-rank partial transfer matrix. The rank depends on the available paths between sources and each intermediate node. More specifically, depending on the topology of the graph, some nodes may receive only combinations of symbols derived from
matrices with restricted rank, i.e.~less than $K$. This includes, for example, trees, where a node connected directly to the source by a link of capacity $C$  can only have children that receive at most rank $C$.

As a first step towards general network models, we consider the case of complete acyclic directed graphs $G=(V,E)$, $n=|V|$, which can
be generated as follows.
\begin{itemize}
\item Generate random labels for the $n$ vertices. These have some ordering $\{e_{1},e_{2},...,e_{n}\}$ associated to them;
\item Make an outgoing (directed) edge from the vertex with the minimum label to every vertex with a higher label;
\item Continue until we reach a vertex where there are no more possibilities for connections.
\end{itemize}

This algorithm generates a complete acyclic directed graph with one source, one sink and $|E|=n(n-1)/2$ edges, since the total degree of each vertex is $n-1= \indegree+\outdegree$. The source and the sink are naturally determined as those nodes that have only outgoing edges or only incoming edges, respectively.
The ordering ensures that this algorithm always generates an acyclic directed graph, conferring the graphs generated in this way specific properties such as the distribution of the in and out-degrees. These properties can be determined directly from the order of the vertex using $\outdegree=n-order(v)$ and $\indegree=n-\outdegree-1=order(v)-1$.

Before proving our next theorem, we introduce the following lemmas.

\lemma{In complete acyclic directed graphs, a node that receives $R$ symbols,  receives w.h.p.~a partial transfer matrix with rank equal to min($R,K$).}\label{prop:deltaDAG2}

\begin{proof} See the {\it Appendix}. \end{proof}

\lemma{ For the complete directed acyclic graph, w.h.p.,
$$\Delta_{S}(v) = \frac{K - \min(K, \textrm{order}(v))}{K}.$$}\label{prop:deltaDAG1}

\begin{proof} See the {\it Appendix}. \end{proof}

\theorem{Let $\phi_{S}$ be the {\em secure max-flow}, defined as the maximum number of symbols that may be secured in a transmission by using random linear network coding. For a complete acyclic directed graph with $n$ nodes, the secure max-flow equals the max-flow min-cut capacity of the network and is $n-1$. Conversely, the minimum numbers of required symbols for secured transmission is $n-1$ symbols.
}

\begin{proof}

Suppose, by contradiction, that $K=n-1$ is the max-flow min-cut capacity of the complete directed acyclic graph. The maximum order of an intermediate node $v$ is $n-2$, thus by \lemmarref{prop:deltaDAG1} we have $\Delta_{S}(v) = 1/(n-1)$.
It follows that the secure max-flow of the complete acyclic directed graph equals the capacity of the graph.

By contradiction, let the minimum number of required symbols for secured transmission be $m_{s}\leq n-2$. There exists an intermediate node $v$ such that $\textrm{order}(v)=n-1$, and consequently, $\Delta_{S}(v) = 0$. Then the minimum number of required symbols for secure transmission is $m_{s}=n-1$.

\end{proof}

It follows that the way to secure this class of complete graphs is to transmit at the max-flow min-cut capacity, if necessary
by adding ``dummy'' symbols.

\section{Conclusions}\label{sect:ConcludingRemarks}

Intrigued by the security potential inherent to random linear network coding,
we developed a specific algebraic security criterion, for which we proved
a set of key properties.
Perhaps one of the most striking conclusions of our analysis is that
algebraic security with network coding is very dependent on the
topology of the network. As an example, we focused on complete acyclic directed graphs,
and determined the secure max-flow, as well as the minimum number of symbols required
for algebraic security. As part of our ongoing work, we are extending this analysis
to other more general network models. Ultimately, we would like to develop secure
communication protocols capable of exploiting random linear network coding as 
an {\it almost free} cypher.

\section*{Acknowledgements}
The authors gratefully acknowledge insightful discussions with Rui A. Costa (Univ. of Porto).

\vspace{-0.2cm}

\appendix

\subsection*{Proof of \lemmarref{prop:rank}}
We will prove this constructively in terms of the ranks of parts of the transfer matrix. The auxiliary encoding vector in each intermediate node $v$ is
given by
$$M'_{\edgesin} = ( A(I-F)^{-1} )_{\edgesin},$$
where $M'_{\edgesin}$ denotes the columns of the matrix corresponding to the incoming edges of $v$. The dimension of  $M'_{\edgesin}$ is $K\times\indegree$, with $\indegree<|E|$.

To determine the rank of the partial transfer matrix, we note that the transfer matrix $M=A(I-F)^{-1}B^T$ for the network must be invertible, and hence, $\textrm{rank}(M)=K$. On the other hand, to determine the rank of $A(I-F)^{-1}$ we use the fact that $(I-F)^{-1}$ is invertible and thus $ \textrm{rank}((I-F)^{-1}) = |E| $. 
We also have $$\textrm{rank}(A(I-F)^{-1}) \leq |E|,$$ because the dimension of $A(I-F)^{-1})$ is $K\times|E|$.
But, since $$ \textrm{rank}(A(I-F)^{-1}B^T) = K = \min(\textrm{rank}(A(I-F)^{-1}),B) $$ holds and $K<|E|$ (true because $K$ must be less than the minimum cut in the network) we conclude that $$\textrm{rank}(A(I-F)^{-1}) = K.$$

We now consider $\Delta_{S}(v)$ at some vertex $v$. For that, we can consider two distinct cases: the first one is if $K < \indegree $. In this case, we cannot assume anything about $\Delta_{S}(v)$, since the rank of the matrix $M'_{\edgesin}$ will be dependent on the topology of the network. As for the second case, $\textrm{rank}(M'_{\edgesin})<K \Rightarrow \Delta_{S}(v)<0$.  
\boxend
\subsection*{Proof of Lemma \ref{prop:linCombFq}}
Contrary to the sum, the product of independent and uniformly distributed values in $\field$ is {\it not} independent and uniformly distributed. In fact, there are two ways to obtain a zero in a multiplication in $\field$: (1) by multiplication between an element $a\in \field$ and $0$, and (2) by multiplication over two elements $a \in \field$ and $b \in \field$, such that $a \neq 0$ and $b \neq 0$, but $ab=0$.
Now, the total number of entries of the multiplicative table between $q$ elements of $\field$ is $q^2$, and
there are at most $2q$ instances of the first case: q instances of $ab=0$, $a=0$ and $b\neq0$, and $q$ instances of $ab=0$, $a=0$ and $b\neq0$. As for the second case, it is possible to prove by contradiction that the number of zeros obtained this way is strictly less than $q^2$: if this was not the case, all products of elements of $\field$ would be zero, and that is absurd. Since this is true for any $q$, the number of zeros grows $O(h(q))<O(q^2)$.
Thus, we have
$$P(X_{lin}=0) \leq \frac{2q+h(q)}{q^2}.$$
Since for large enough $q$ we have $(2+h(q))/q < 1$, it follows that
$$P(X_{lin}=0)_{q\rightarrow\infty} = 0.$$ \boxend
\vspace{-0.25cm}
\subsection*{Proof of Lemma \ref{prop:lineTransferMatrix}}
Each position of a line of the transfer matrix $M$ is a linear combination of independently and uniformly chosen values in $\field$, and thus, the probability of obtaining a zero in a position is given by \lemmarref{prop:linCombFq}. The result follows by considering all the combinations of the possible positions in which the $Y$ zeros may occur. \boxend

\vspace{-0.25cm}
\subsection*{Proof of Lemma \ref{prop:deltaDAG2}}

Suppose that a given intermediate node receives $R=K+\theta$ symbols, $\theta\ge0$.
It is clear that the maximum possible rank is $K$ and thus there is a way to remove $\theta$ columns s.t. the rank of the resulting set will still be at maximum $K$.
Now consider the case in which vertex $v$ receives at most $K$ symbols. If the columns are linearly dependent, the condition
$$ \{x_{h_{1}}\underline{c}_{h_{1}}+x_{h_{2}}\underline{c}_{h_{2}}+...+x_{h_{n}}\underline{c}_{h_{n}}=(0...0)^T\},$$
such that $x_{h_{1}}, x_{h_{2}}, ..., x_{h_{n}} \textrm{not all } 0, \in \field$ and $h_{1},h_{2},...,h_{n}$ represent the columns  $\in \edgesin$, will be satisfied. Since the linear combination of lines of the transfer matrix is again a linear combination of independent and uniformly distributed values in $\field$, it follows from \lemmarref{prop:lineTransferMatrix} that the probability of obtaining $(0...0)^T$ tends to $0$ when $q\rightarrow\infty$ and $K\rightarrow\infty$, and thus, the columns $h_{1},h_{2},...,h_{n} \in \edgesin$ are linearly independent w.h.p.
\boxend\vspace{-0.25cm}
\subsection*{Proof of Lemma \ref{prop:deltaDAG1}}

It follows from \lemmarref{prop:deltaDAG2} that w.h.p., the number of symbols received by a vertex is the rank of the partial transfer matrix received (and at most $K$) and thus
\begin{align*}
\Delta_{S}(v) =
&\frac{K - \min(K,\indegree)}{K} =&\\
&\frac{K - \min(K, \textrm{order}(v)-1)}{K}
\end{align*}
\boxend

\end{document}